\documentclass{emulateapj}                                            %

\newcommand{\be}{\begin{equation}}
\newcommand{\ee}{\end{equation}}
\newcommand{\ba}{\begin{eqnarray}}
\newcommand{\ea}{\end{eqnarray}}

\newcommand{\saxj}{\mbox{SAX~J1808.4-3658 }}
\newcommand{\herx}{\mbox{Her~X-1 }}

\newcommand{\km}{\hbox{km}}

\slugcomment{Accepted by ApJ  Sept. 16, 2007}
\shorttitle{Mass and Radius Constraints}

\begin{document}

\title{Limits on Mass and Radius for the ms-Period X-ray Pulsar SAX~J1808.4-3658}
\author {Denis A. Leahy\altaffilmark{1}, Sharon M. Morsink\altaffilmark{2,3},
Coire Cadeau\altaffilmark{2}}

\altaffiltext{1}{Department of Physics and Astronomy, University of Calgary,
Calgary AB, T2N~1N4, Canada}
\altaffiltext{2}{Department of Physics,
University of Alberta, Edmonton, AB, T6G~2G7, Canada}
\altaffiltext{3}{On sabbatical at: Pacific Institute for Theoretical Physics,
Department of Physics and Astronomy, University of British Columbia,
6224 Agricultural Road, Vancouver  BC, V6T~1Z1, Canada}

\begin{abstract}
\saxj has a 2.5 millisecond neutron star rotation period and exhibits X-ray 
pulsations due to its rotating hot spot. Here we 
present an analysis of the pulse shapes of \saxj during its 1998 outburst.
The modeling of the pulse shape includes several effects, including gravitational 
light-bending, doppler effects and two spectral components with different emissivity.
In addition we include the new effects of light-travel time-delays and
the neutron star's oblate shape. We also consider two different data sets, with
different selection in time period (1 day versus 19 days of data combined) and different
energy binning and time resolution. We find that including time-delays and oblateness
results in stronger restriction on allowed masses and radii. A second result is that
the choice of data selection strongly affects the allowed masses and radii. Overall,
the derived constraints on mass and radius favor compact stars and a soft equation of state.
\end{abstract}


\section{Introduction}
\label{s:intro}

The discovery  \citep{WvdK98}  of 2.5 ms pulsations originating from \saxj
(hereafter SAX J1808)
provides strong evidence that the neutron stars in low mass X-ray
binaries are the progenitors of the millisecond period pulsars. SAX J1808
is now one of 7 known accreting ms X-ray pulsars (see \citet{Wij05}
for an observational review of the properties of the first 6 pulsars and
\cite{Po06} for an updated review including the first 7 pulsars). 

The pulsed X-rays observed during outburst are most likely produced 
from the energy released from accretion of plasma funnelled onto the
neutron star's magnetic poles (see for example, Figure 12 of 
\citet{GDB02}). Spectral models \citep{Gil98,GDB02} provide
strong evidence that the X-rays correspond to blackbody emission from
a spot on the star which is then Compton scattered by
electrons above the hot spot. Since, in this model,
the pulsed light is emitted from the neutron star's surface (or from a 
region very close to the surface) the accreting ms X-ray pulsars 
are excellent targets for light curve fitting in order to constrain
the neutron star equation of state. The X-ray light curve depends on
the intrinsic properties of the emission (spot shape, size, location
and emissivity) as well as the neutron star's macroscopic properties
(mass, radius and spin). If tight enough constraints on the star's 
mass and radius can be made, it could be possible to constrain the
equation of state of super-nuclear density material.

The first pulse shape analysis \citep{PG03} (PG03) for SAX J1808 provided 
interesting constraints on the neutron star's mass and radius. 
However, this analysis did not take into account two effects 
that are potentially important for rapidly rotating neutron 
stars: variable time delays due to light travel
time across the star \citep{CLM05} and the oblate shape of the
star \citep{CMLC07}. One of the motivations for the re-examination
of the pulse shapes for SAX J1808 is to include these effects in
the analysis. In this paper we re-analyze this data in order to
determine the importance of these effects. In order to isolate 
these effects, we have kept all other aspects of our data
analysis as close as possible to that of PG03.

Additional aspects that we explore are: a) the effects of time and energy binning
of the data on the fitted parameter values; and b) the effect of data selection. 
\citet{Pap05} constructed a light curve for SAX J1808 using an subset of the data
used by PG03 and also used a different binning in energy and time. 
Since SAX J1808 was quite variable during the time period analyzed by PG03 (see Fig.1 and
Fig.2 of \citet{GDB02}), it is important to carry out an analysis on data
set from a shorter, less variable time interval.  
Thus we present
independent fits to the \citet{Pap05} light curve in order to 
explore the sensitivity of the fits to different types of binning and to choice of data
interval.

The outline of our paper is as follows. In section~\ref{s:method}
we explain in detail the method and types of models that we use to analyze the data. 
In section~\ref{s:results} we present the results of our fits for
light curves separated into two energy bands and show the effects due
to changes in spectral models, time-delays and the star's oblateness.
In section~\ref{sec:bolo} we present the results of our fits for
a bolometric light curve.
We conclude with a discussion
of the mass-radius constraints found in our analysis.

\section{Method}
\label{s:method}

SAX J1808 went into outburst in April 1998. From the RXTE observations
of the 1998 outburst, two groups have constructed light curves. We now 
describe the different types of binning that were done by each group
in order to construct the light curves.

\begin{figure}
\plotone{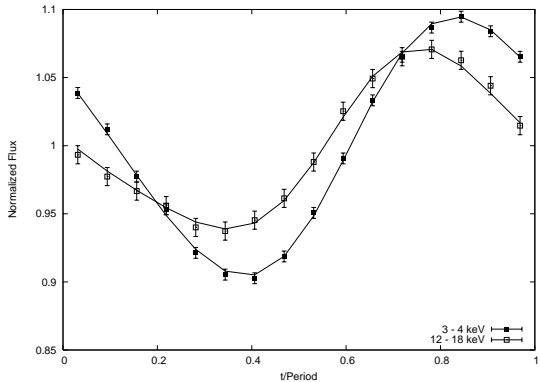}
\caption{Two-band data from the 1998 outburst for SAX J1808 and the best-fit model. 
Error bars are 2 $\sigma$.
Data is reproduced from \citet{PG03}.
Best-fit model with $2M/R=0.6$ (from Table~\ref{tab:ob}) is
shown as solid curves.
}
\label{fig:PGdat}
\end{figure}

The two light curves constructed by PG03 are shown in their Figure 5a
and reproduced in Figure~\ref{fig:PGdat}.
In their analysis, they constructed light curves in two energy
bands, a low energy band (3-4 keV) and a high energy band (12-18 keV). 
For each band they 
folded  data from April 11 - 29 into light curves with 16 time
bins per period. We refer to the light curves constructed by
PG03 as ``two-band light curves''.

\begin{figure}
\plotone{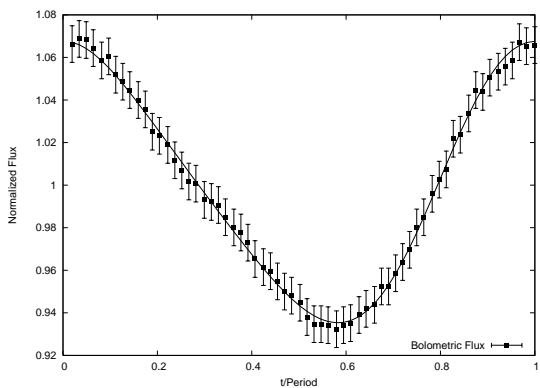}
\caption{Bolometric data from the 1998 outburst for SAX J1808 and the best-fit model. 
Error bars are 2 $\sigma$.
Data is reproduced from \citet{Pap05}.
Best-fit model with $2M/R=0.6$ (from Table~\ref{tab:Pap-ob}) is
shown as a solid curve.
}
\label{fig:Papdat}
\end{figure}

\citet{Pap05} constructed a light curve shown in
the upper panel of their Figure 2 and reproduced here as
Figure~\ref{fig:Papdat}. This light curve was constructed
by combining the data from all energies in the range 2 keV - 60
keV collected during an approximately 22 hour period starting 
on April 18. This data is folded into one light curve with
64 time bins per period. The error bars for the bolometric data
is about 2 times larger than the two-band data due to fewer counts per 
bin: partly due to more bins (64 compared to a total of 32 for the PG03
light curves) and partly due to a shorter time period for data selection. 
We refer to the light curve 
constructed by \citet{Pap05} as a bolometric light curve.

The data used in the two types of light curves overlap to a significant
extent in energy and in time period selected. Thus
we expect that the conclusions deduced independently from
each binning should be roughly consistent with each other. The two main differences
between the two methods are: a) a shorter time period selection for the \cite{Pap05}
light curve; and b) the binning by energy. The 
two-band light curves constructed by PG03 make use of one 
narrow energy band and one broad energy band 
and do not use the data from 5 to 12 keV or above 
18 keV. The bolometric light curve constructed by \citet{Pap05} 
makes use of all the energy channels, but does not provide
information about the change in pulse shape with energy.
However, the \citet{Pap05} light curve provides 4 times better time 
resolution. In our modelling we will fit the data sets independently.

\subsection{Two-Band Light Curves}
\label{s:spectrum}

Spectral models by \citet{GDB02} show that the spectrum
is well approximated by the sum of a blackbody 
and a Comptonized spectrum. In the low energy band, the 
blackbody flux is about 30\% of the Comptonized flux,
while in the high energy band the blackbody radiation 
is negligible. In our models for the two-band
light curves, we use three spectral
components to model the data. These components are
(1)~Comptonized flux in the high energy band,
(2)~Comptonized flux in the low energy band,
and (3) blackbody flux in the low energy band. 
Models for each spectral component in the star's rest frame
(see Figure 3 of \citet{GDB02}) follow a power-law form
if only the limited range of energies for the appropriate
energy band are considered. For each component, we model 
emitted monochromatic flux with the following function
\be
F_{i,em}(E_{em},\mu) = A_i(\mu) E_{em}^{-\Gamma_i + 1}
\label{Fem}
\ee
where $E_{em}$ is the energy in the ``emission'' frame
that rotates with the star, and $\mu$ is the cosine
of the angle between the normal to the star's surface and
the initial photon direction, as measured in the 
emission frame. The subscript ``i'' takes on values 
of 1,2 or 3 corresponding to the three spectral 
components. The functions $A_i(\mu)$ describe the anisotropy
of the emitted light.

In the observer's frame, Doppler effects must be 
taken into account in order to find the observed flux.
The transformation laws have been described in detail
elsewhere (see PG03, \citet{CMLC07}) and will be briefly summarized 
here. The Doppler boost factor is defined by 
$\eta = (1-v^2)^{1/2}/(1-v\cos\xi)$ where $\xi$ is
the angle between the fluid's velocity vector and the
initial photon direction and $v$ is the magnitude
of the fluid velocity at the latitude $\theta$ 
of emission, $v=\Omega_* R (1-2M/R)^{-1/2} \sin \theta$,
and $\Omega_*$ is the star's angular velocity.
The observed energy is given by $E_{obs}=\eta E_{em}$,
the specific intensity transforms as $I_{obs}(E_{obs}) = \eta^3I_{em}(E_{em})$
and the solid angle subtended by the spot transforms as
$d\Omega_{obs} = \eta d\Omega_{em}$. Since the flux
is given by $F(E) = I d\Omega_{obs}$, the observed flux at energy $E_{obs}$ 
for spectral component $i$ is given by
\be
F_{i,obs}(E_{obs}) = \eta^4 F_{i,em}(E_{em}).
\label{Fobs}
\ee
For the powerlaw components given by equation (\ref{Fem}),
the observed flux for each component is
\be
F_{i,obs}(E_{obs}) = \eta^{3+\Gamma_i} A_i(\mu) E_{obs}^{-\Gamma_i+1}.
\ee
The Doppler boost factor and $\mu$ depend on phase, so the monochromatic
flux also depends on phase. If the monochromatic fluxes are integrated 
over the appropriate observed energy ranges the phase dependent factors
are unchanged, and the integrated flux for each component
is
\be
F_{i} = I_i \eta^{3+\Gamma_i} A_i.
\ee

The quantity measured in an X-ray detector is not flux, but photon
number counts, $N(E)$ for a specified energy. The monochromatic flux
is related to the photon number counts by $F(E) = E N(E)$. Since the 
photon number counts in the emission frame (for a given component) are given by 
$N_{i,em}(E_{em}) = E_{em}^{-\Gamma_i}$ arguments similar to those given above show
that the observed integrated photon number counts have the same phase dependence
as the observed integrated flux, $N_{i,obs} \propto \eta^{3+\Gamma_i} A_i(\mu)$. 
Hence it does not matter whether flux or photon number counts are used.

Our spectral model uses fixed values of $\Gamma_i$ determined by
the best-fit model spectrum used by PG03 (see their Figure 3). The values of
$\Gamma_i$ are $\Gamma_1=2.0$ (high energy band and Compton),
$\Gamma_2=1.44$ (low energy band and Compton), and $\Gamma_3=3.34$ 
(low energy band and black body).  
The anisotropy function for the Compton components in both energy
bands are assumed to have the form
\be
A_{1,2} = 1 - a \mu
\label{eq:aniso}
\ee
where we assume that the beaming is independent of energy. 
This simple parameterization has been shown by PG03 to be a 
reasonable approximation to the calculations by 
\citet{Sun85}. For the blackbody component, we assume that the
emitted light is isotropic. In our initial models we included
an exponential absorption factor ($~e^{-\tau/\mu}$) but we 
found that including a non-zero optical depth $\tau$ did 
not significantly affect our fits, so we set $\tau=0$. 
The constants $I_1$ and $I_2$ are free parameters in 
our models. The constant $I_3$ is defined after introducing
the parameter $b$, the ratio of the phase averaged 
blackbody flux to the phase averaged Comptonized flux in the
low energy band.
The constant $b$ is defined by $b = \bar{F_3}/\bar{F_2}$
where the bar refers to an average over phase. In our
fits we restrict $b$ to within 15\% of the 
value based on the PG03 spectral model.


\subsection{Bolometric Light Curve}

The bolometric light curve is modeled using a method similar to
the two-band light curves. The Compton component is modelled as
a powerlaw with photon spectral index $\Gamma=2$ over the entire energy range, and 
the anisotropy is modeled using the one-parameter function given
by equation (\ref{eq:aniso}). The blackbody component is modelled 
as a powerlaw with photon spectral index $\Gamma=3.34$ and is normalized
so that the blackbody to Compton ratio in 
the 2-60 keV band agrees with the PG03 spectral model.

\section{Results for Two-Band Light Curves}
\label{s:results}

In this section we apply a series of different methods to compute
models for the two-band light curves constructed by PG03. Our main
goal is to test the importance of including time-delays and oblateness
in the analysis of data from SAX J1808. In order to test these effects
we begin by reproducing the original fits of PG03 and then adding
sequentially the different effects.

\subsection{Fiducial Models}
\label{s:fiducial}

We begin by reproducing the analysis of PG03 as closely as possible.
To do this we use the spectral model described in section
\ref{s:spectrum}. We keep the spot size fixed so that the radius in
the star's frame is 1.5 km. (Later we show that changes in the spot size
do not significantly change the fits.) For the first set of fits
(which we refer to as ``fiducial''), the exact light-bending angle
formula is used, time delays are omitted and the surface of the 
star is assumed to be spherical. The results for the fiducial
best-fit models are shown in  Table \ref{fid}.

For our fits we fix the value $2M/R$ in order to simplify the
computation of the bending angles and the time delays. In order to
correctly include the relativistic bending of light rays we require
a numerical solution for the relationship between $\alpha$, the angle
between the initial photon direction and the radial vector pointing to
the spot,
and $\psi$, the angle between the final photon direction and the
spot's radial vector. This relationship requires solving the integral
relationship given by \citet{PFC83} and depends on $2M/R$. Similarly,
the time-delays also depend on the value $2M/R$. Given a value of
$2M/R$, it is then possible to interpolate from the same numerically
generated table relating $\alpha$ and $\psi$ for each trial value of
neutron star mass. We note that PG03 did their fits by first fixing a
value of mass and then allowing $2M/R$ to vary.

 For each row of Table
\ref{fid}, the ratio of $2M/R$ is fixed and the other
parameters are allowed to vary. The 8 free parameters are:
$M$, the mass of the neutron star; $\theta$, the co-latitude of the
spot's centre; $i$, the inclination angle of the observer;
the parameters $I_1$, $I_2$, $b$, and $a$ describing the spectral
model; and $\phi$ a phase angle. Since there are 32 data points
and there are 8 free parameters, the fits have 24 degrees of freedom. 
Our best-fit parameter values are
slightly different from those of PG03, and our lowest $\chi^2$ is
larger than that of PG03. However, our best-fit models still agree 
with the best-fit models of PG03. 

In our models we do not restrict the values of the inclination
angle for the system, although distance-dependent limits on $i$ 
have been derived by \citet{CM98} and \citet{Wang}. The 
more recent distance determination by \citet{GC06} suggests
a small inclination angle, but since there is still a fairly
large uncertainty in $i$, we choose to keep its value free.

We now compare the results of our fiducial fits with those of
PG03, given in their Table~1 and labelled ``Model~2''. First,
consider the model with the lowest value of $\chi^2$ in 
Table~\ref{fid}, corresponding to $2M/R = 0.6$. 
This model is very similar to the best fit model given in Table~1
of PG03 which has $M=1.0 M_\odot$, $R=5.0$ km and $2M/R = 0.599$. 
Our best-fit model has a radius that is about 10\% larger than
the PG03 best-fit model. The best-fit inclination angle, spot 
co-latitude and anisotropy parameter $a$ are all 
smaller in our best-fit model compared to the 
PG03 best-fit model, but the differences are within the error
limits given by PG03. 

It should be remembered that neutron stars
with $2M/R \ge 0.57$ have regions where light can be emitted in two
different directions and still reach the observer. However, this
region is very small for the case of $2M/R=0.6$, and we have checked 
that in our best-fit solution the spot never enters the region where
multiple images would occur. The largest value of this ratio used
in our computations is $2M/R=0.6$ since properly taking multiple images 
into account is computationally difficult.
Since the light curves are very close to
sinusoidal, it would be very unlikely for a geometry producing
multiple images to fit the data. 

As the mass of the star is increased, PG03 find that the 
best-fit $\chi^2$ increases rapidly. In their models, the
$1.4 M_\odot$ model has a $\delta\chi^2=10$ compared to their
model with $1.0 M_\odot$. This means that the $1.4 M_\odot$ is
a significantly worse fit (by about 3 $\sigma$) than the fit
for the $1.0 M_\odot$ star. We attempted to fit the data with 
$1.4 M_\odot$ stars while allowing other parameters to vary.
The resulting minimum value of $\chi^2=40.3$ for a $1.4 M_\odot$ star
(with a radius of 8.3 km)
is larger than our best-fit vale of $\chi^2=31.1$ obtained with M as free parameter:
a similar result to that obtained by PG03.

The reason why only very small radius (and small mass) stars are allowed by the
data can be understood by examining the low-energy band light curve
reproduced in Figure~\ref{fig:PGdat}. The low-energy light curve
has a shape that is very close to sinusoidal. The main effect of the
Doppler factor $\eta$ is to increase the asymmetry in the light curve.
The maximum value of $\eta$ is related to the speed of the
fluid at the emission region, which is proportional to
$R \sin\theta\sin i$. In the low energy band the model spectrum
includes a blackbody spectrum with a large effective slope
($\Gamma=3.34$). Since the flux varies as $\eta^{3+\Gamma}$,
a large speed would create a large asymmetry in the pulse shape. Hence
in order to fit the almost sinusoidal shape of the low-energy band
curve, a very small radius is required. It could be argued
that a large radius could be allowed if the angles $\theta$ and/or
$i$ are small enough. However, PG03 derived a simple approximation for
the variability amplitude which scales roughly as $\sin\theta\sin i$
for small values of the angles. If these angles are chosen to be too
small, then the resulting variability will be too small to match the
light curve.

\begin{figure}
\plotone{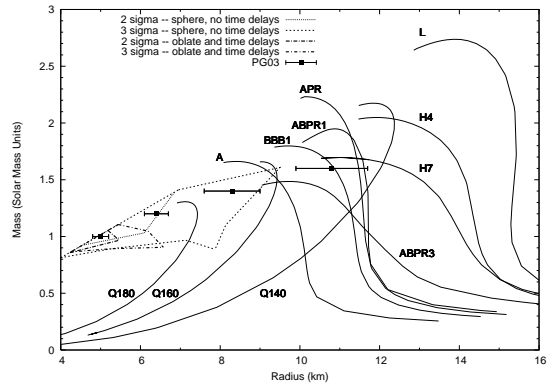}
\caption{Mass and radius confidence contours for two-band data. 
Confidence contours corresponding to $2 \sigma$ (95.4\%) and $3
  \sigma$  (99.7\%) for models assuming a spherical surface and no
  time delays are plotted as dashed curves. 
Confidence contours corresponding to $2 \sigma$ and $3 \sigma$ 
for models assuming an oblate surface and including time delays
are shown as bold dot-dashed curves. 
Best fit models computed by PG03 are shown as squares with errorbars.
Mass versus radius curves for compact stars rotating at 401
  Hz are shown as solid curves. 
The EOS plotted are: A \citep{AB77}, ABPR1-3 (mixed quark and hadron
  phase) \citep{ABPR}, APR \citep{APR}, BBB1 \citep{BBB},
H4 \& H7 (hyperons) \citep{LNO}, L \citep{AB77},
and quark stars denoted Q140-Q180 (where the number corresponds to the value
of $B^{1/4}$ in Mev and $B$ is the bag constant) \citep{Gle00}.
}
\label{fig:mr1}
\end{figure}

In Figure \ref{fig:mr1} we show mass-radius curves for neutron 
stars spinning at 401 Hz constructed with a variety of different
equations of state. On Figure \ref{fig:mr1} we plot the $2 \sigma$ 
and $3 \sigma$ confidence contours arising from the fits assuming
a spherical surface and no time delays as dashed curves. The 
contours are found by varying all 8 free parameters and finding
contours of constant $\chi^2$ corresponding to values $\delta \chi^2 =
4$ ($2 \sigma$) and $\delta \chi^2 = 9$ ($3 \sigma$) larger than the
global minimum of $\chi^2_{min}=31.1$ found for the $2M/R=0.6$ model.
We also include for the purpose of
comparison the best-fit stellar models 
found by PG03 including their 99\% error bars. However, the PG03 
error bars are for stars with a fixed value of mass whereas our
confidence contours allow mass to be a free parameter.
The $3 \sigma$ confidence region shown in Figure~\ref{fig:mr1}
includes a model with a mass as high as $1.6 M_\odot$ with 
a radius of 9.5 km, consistent with both quark stars and
the soft neutron star EOS A.


\subsection{Approximate Light-Bending Formula}

\citet{Bel02} derived a simple approximate formula for light-bending,
\be
\cos\alpha \sim \frac{2M}{R} + \left(1 - \frac{2M}{R}\right) \cos
\psi.
\label{eq:bel}
\ee
The approximate equation (\ref{eq:bel}) is least accurate for light
emitted close to the tangent to the star's surface. Since this formula is
very simple it is interesting to see how the star's parameters
extracted from light curve fitting using the approximate light-bending
formula compare to the case of exact light-bending. In 
Table~\ref{tab:approx} we show the results of fits using a method
identical to the method used for the fiducial models except that the
approximate light-bending formula (\ref{eq:bel}) is used. 
The resulting best-fit models have radii that differ by less
than 10\% than the best-fit model found using the exact
light-bending formula. The main reason why the approximate formula
works in this case is that the light is never emitted very close to
the tangent. For stars with much more modulated light curves we might
expect a larger error. For the rest of the paper we return to the use
of the exact light-bending formula.

\subsection{Uncertainty in the Spectral Model}

There is some uncertainty in the underlying spectral model for the
emitted light. The spectrum is measured in the inertial frame, and in
order to infer the spectrum in the frame moving with the star, some
assumptions about the geometry of the spot and the observer 
and the distance to the source must be made. 

In order to test the dependence of the fitted parameters on the
details of the spectral model we now vary some of the parameters that
are fixed in our fiducial model. For the following cases we fix 
$2M/R=0.6$ and present the results in Table~\ref{tab:spectra}. For
each row in Table~\ref{tab:spectra} we provide the parameter that has
been changed from its value in the fiducial model. For reference the
first column of Table~\ref{tab:spectra} is the fiducial model
shown in Table~\ref{fid}.

The simplest spectral model is a powerlaw with a fixed value of
$\Gamma$ for all components. \citet{GDB02} showed that the
best fit for the photon spectral index is $\Gamma=1.8$,
although their Figure 2 shows $\Gamma$ on individual days varying between 1.9
and 2.1.
 In the row labelled $\Gamma=1.8$ in 
Table~\ref{tab:spectra} we show the results of fits with the spectral
photon index fixed at this value for both the blackbody and Comptonized 
components.  The resulting best fit model has parameters that are very close
to the fiducial model, but with a larger value of $\chi^2$. 

In the spectral models used by PG03, an absorption factor was included for
the blackbody component with the form $\exp(- \tau/\mu)$. In the row labelled
$\tau=0.1$ we consider the effect of nonzero $\tau$. For this test we fix
$\tau$ to a value consistent with the best fit values found by PG03. 
The result is a better fit, at low significance (1.2$\sigma$), but the 
best-fit parameters are within a few
percent  of the fiducial values (with $\tau=0$).

In all of the fits so far we have kept the ratio $b$ of the average blackbody
to Compton flux in the low-energy band fixed to within 15\% of the
model used by PG03. We now allow this ratio to be free while keeping
the values of the photon spectral indices the same as in the fiducial
model.  The results of this fit is shown in the row 
labelled ``$b$ free'' of Table~\ref{tab:spectra}. Allowing $b$ to be free
gives the most significant decrease in $\chi^2$. However the changes in the
best-fit parameters are again very small.

In the spectral model used by PG03, the radius of the spot as measured
at infinity was fixed at $r_{\infty} = 2.4$ km, corresponding to a spot
size on the star ranging from 3.0 km to 3.7 km depending on the 
star's assumed mass. In the case of the model with $2M/R=0.6$ 
the spot size on the star is 3.0 km. For rows labelled 
$r_{sp} = 1\km - 3\km$ in Table~\ref{tab:spectra} 
we now allow for different values of the spot radius. (For reference, the spot size for the fiducial model
is 1.5 km)
 Allowing for a larger spot (3.0 km) increases the radius by  3\%
and decreases 13 the observer's inclination angle by 13 degrees while 
increasing $\chi^2$.  The 
dependence of the inclination angle on  the spot size suggests that the 
best-fit inclination angle should not be considered to be accurate.

\subsection{Time Delays}
\label{sec:td}

Two photons emitted simultaneously from the front and back of the star 
will arrive at the detector at different times separated by an
interval of order $\Delta t \sim 2R/c$. If the star has spin period
$P$ and the data is binned into $N$ bins per period, then when the 
dimensionless ratio, $\kappa$, of the time delay to the time per bin
\be
\kappa = \frac{2 R N}{c P}
\ee
is of order unity then it becomes important to correctly bin the
photons based on the variable times of arrival. For SAX J1808,
$\kappa \sim 0.025 N (R/10 \km)$, so for 16 time bins (as in the PG03
data) and a 10 km star this ratio is $\kappa \sim  0.4$, so we expect
 that the variable time delays may be important for the larger
stars. For the light curves constructed by \citet{Pap05} with 64 time
bins, $\kappa \sim 1.7$ and the inclusion of the time delays is
crucial in order that the light curve be modeled correctly. 

When the variable time delays are included in a model light curve, the
time delays tend to increase the asymmetry of the resulting light
curve (compared to a light curve where the time delays are ignored). 
Since the addition of variable time delays creates an extra asymmetry,
the asymmetry created by the Doppler boost factors must be reduced,
resulting in a smaller radius. Hence, we expect that the addition of
time delays to the light curves we result in a reduction in the size
of the best fit stars, as discussed by \citet{CLM05}.

In Table~\ref{tab:td} the best fit models including time delays are
shown. All other aspects of the models are identical to the fiducial
models computed in section~\ref{s:fiducial}. Comparing
Tables~\ref{fid} and \ref{tab:td} we see that for fixed values of 
$2M/R$ the radius (and mass) decreases by about 10\% when time delays
are included.

The most important effect of including time delays is a large
increase in $\chi^2$ (for fixed values of $2M/R$) that can be 
seen by comparing Tables~\ref{fid} and \ref{tab:td}. This 
effect occurs because the inclusion of time delays in the model
changes the shape of the light curves, making it harder to fit the
data. This increase in $\chi^2$ has the effect of narrowing
the allowed values of $2M/R$ to a smaller range.

\subsection{Oblate Shape of Star}

A rotating star has an oblate shape. The most important effect that
the oblate shape has on the light curves is the change in the
visibility condition \citep{CMLC07,MLCB07} for photons. The effect on
light curves is most pronounced for large stars, since they are more
oblate. However, if the emission-observer geometry is such that some
of the photons must be emitted close to the tangent to the star, then
the light curves constructed from oblate and spherical stars can be
very different. 

In cases where both the observer and the spot are located in the same
hemisphere (as defined by the star's spin equator), 
our previous calculations \citep{CMLC07} have shown that
if the inclination angle, spot latitude and the ratio $2M/R$ at the
spot's location are kept fixed, the light curve constructed from an
oblate star will be less modulated than the light curve constructed
from a spherical star. The oblate shape makes it easier to see the
spot when it is at the back of the star. 
The opposite effect occurs if spot and observer are located in
opposite hemispheres.

We have developed a simple approximation \citep{MLCB07} that captures
the essential features of oblate stars. In this approximation, we find
that the deflection angle and times of arrival are approximated very
well by the Schwarzschild metric. However initial conditions for the
directions that the photons can be emitted into must take the oblate
shape into account. We have found a simple formula 
\citep{MLCB07} for the oblate
shape that depends only on the ratios $M/R$ and $\Omega^2R^3/M$
and has very little equation of state dependence. 

Since the inclusion of an oblate shape decreases the modulation of a
light curve, we can make some predictions about the effect of oblateness on the fitted
radius. Suppose that the ratio $M/R$ is kept fixed, as is done in our
fitting procedure. Adding oblateness while keeping $i$ and $\theta$
fixed will result in a light curve that is not modulated enough. In
order to fit the modulation correctly a fit with $i+\theta$ larger
than in the spherical case will be preferred. (This is because we need
to increase the angular separation of the spot and the observer when
the spot is at the back of the star.) An increase in $i+\theta$ will
change the magnitude of the quantity $R\sin\theta\sin i$ which 
controls the asymmetry of the light curve. Since 
$\sin\theta \sin i = ( \cos(\theta-i) - \cos(\theta+i))/2$ there are
two possible cases when $\theta+i$ increases. In the first case,
the quantity $|\theta-i|$ decreases, which leads to an overall 
increase in the quantity $\sin \theta \sin i$. In order to keep the
asymmetry the same, the best-fit star must have a smaller value of $R$ 
to compensate. In the second case, the quantity $|\theta-i|$ increases,
in which case the change in $\sin\theta \sin i$ is indefinite
and the star's radius could either increase or decrease
when oblateness is added. Although it is possible in principle for either case to
occur, we have found that only the first case occurs for the 
models computed in this paper.

The general trend in the best-fit angles and stellar radius can be seen by comparing the results 
of the oblate fits in Table~\ref{tab:ob} with the results of the fits
for spherical stars in Table~\ref{tab:td}. In each case (of fixed
$M/R$) the addition of oblateness increases the combination of angles
$\theta+i$ while decreasing the magnitude of their difference
$|\theta-i|$, leading to a decrease in the best-fit radius as
given in the first case as described in the previous paragraph.
However, only in the
case of the largest stars does the decrease in radius 
due to oblateness rival
the decrease in radius due to the  inclusion of time delays.

We again attempted to fit the data with 
$1.4 M_\odot$ stars while allowing all other parameters to vary using the
oblate models with time delays.
The resulting minimum value of $\chi^2=237$ for a $1.4 M_\odot$ star
is unacceptably large and strongly excludes larger mass stars.

In Figure \ref{fig:mr1} we plot the $2 \sigma$ 
and $3 \sigma$ confidence contours arising from the fits including
time delays and oblateness as bold dot-dashed curves. The inclusion
of time delays and oblateness shrinks the allowed values of mass
and radius to a very small region of Figure~\ref{fig:mr1} compared to
the allowed region when time delays and oblateness are not included.
The shrinkage of the allowed region is mainly due to the increase
in $\chi^2$ introduced when the time delays are included, as
noted in section~\ref{sec:td}.

The allowed region of the mass-radius plane at the 3 $\sigma$ 
confidence level only allows compact stars with very small radius
and mass when time-delays and oblateness are included. The largest
radius star in this region has a radius of 6.6 km and a mass of 
$0.9 M_\odot$. The largest mass star in this region has
a radius of 5.4 km and a mass of $1.1 M_\odot$. These values are
inconsistent with any known neutron star EOS, but could be 
described by a quark star EOS if the bag constant is
larger than usually considered ($B^{1/4} \sim 200$ MeV).

\section{Bolometric Light Curve}
\label{sec:bolo}

The bolometric light curve constructed by \citet{Pap05} presents an
alternative method for binning the data than that used by PG03. There 
are advantages and disadvantages to both methods. The PG03 method
has the advantage of explicitly separating the effects of the different 
contributions from blackbody and Comptonized components. The bolometric light
curve has the advantage of including photons of all energies and also has four
times better time resolution, which is important when modeling the asymmetry
in the light curve. The time period
selected by \citet{Pap05} is much shorter, which has the advantage of avoiding
mixing time periods from SAX J1808 where the pulse shape is different due to 
variability in the parameters of the emission region.

In our models of the bolometric light curve, the ratio $2M/R$ is kept
fixed as in the case of the two-band light curves. The free parameters
are: $M$, $I$, $\theta$, $i$, $b$ and $\phi$. The parameter $I$ is the
normalization of the Comptonized component of the flux. Similar to the definition
earlier in section~\ref{s:spectrum} the parameter $b$ is the ratio of the
blackbody flux to the Componized flux in the energy range 2 - 60 keV
and is allowed to vary within 15\% of the PG03 spectral model. As in the
two-band models, the spot radius is kept fixed at 1.5 km. 
Since
there are 64 time bins, the degrees of freedom in the bolometric fits
are 58. 

\subsection{No Time-delays and Spherical Surface}

In Table~\ref{tab:Pap-notd} we show the results of fitting the bolometric 
light-curve using a model that assumes a spherical surface 
and omits time delays. This Table should be compared to the results of the 
fiducial two-band results shown in Table~\ref{fid}. In the case of the
bolometric data, there is very little variation in the minimum value 
of $\chi^2$ for fixed values of $2M/R$ in contrast to the two-band data
which favours large values of $2M/R$. As a result, the bolometric
light curve is consistent with a larger range of masses and radii than
allowed by the two-band light curves.

\begin{figure}
\plotone{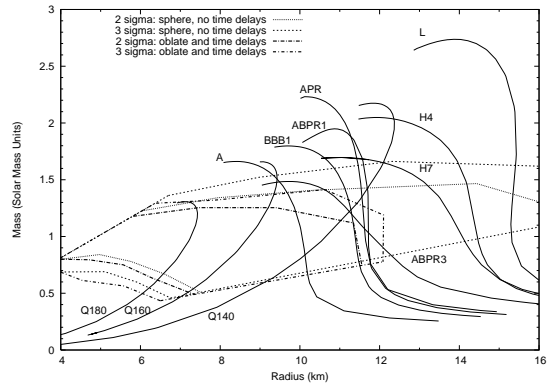}
\caption{Mass and radius confidence contours for bolometric data.
Confidence contours 
corresponding to $2 \sigma$ and $3  \sigma$ for models assuming
a spherical surface and no time-delays are plotted as dotted
curves. Confidence contours for models assuming an oblate surface
and including time-delays  are shown as bold dot-dashed curves.
Mass versus radius curves for compact stars rotating at 401
  Hz are also shown as solid curves.
The EOS labels are identical to those used in Figure~3. 
}
\label{fig:mr2}
\end{figure}

The region of the mass-radius plane allowed at the 3 $\sigma$ 
confidence level is shown in Figure~\ref{fig:mr2} as a dotted curve.
This allowed region includes the corresponding region allowed by
the two-band fits. However, the region consistent with the bolometric
light curve includes a much larger range of masses and radii than
the two-band light curves allow. When time-delays are omitted and a
spherical surface is used, the bolometric data is consistent with very
stiff EOS such as those including hyperons.


\subsection{Model Dependence of Bolometric Data}

We now test the dependence  of the fits to the bolometric light curve 
on the assumed model. In Table~\ref{tab:Pap-models} we show the results
of these tests, where the ratio $2M/R=0.6$ is kept fixed.
All models are computed using exact light-bending,
no time delays and a spherical surface. The first row of Table~\ref{tab:Pap-models}
is repeated from Table~\ref{tab:Pap-notd} and in each subsequent row one parameter
in the model is changed. The rows labeled $r_{sp}=1-3 \km$ show the dependence of the
best-fit parameters on the assumed spot size. The best-fit stellar radius varies 
by less than 5\% as the spot size is changed, similar to the dependence on spot size
seen in the two-band fits. The row labeled $\Gamma=1.8$ shows that the star's radius
increases by 3\% when the photon spectral index is decreased from 2.0 to 1.8.

\subsection{Time-Delays and Oblateness}

In Table~\ref{tab:Pap-ob} we show the results of fitting the
bolometric data when we include the effects of variable time delays
and the oblate shape of the star in the theoretical models. As in the
case of the two-band light curve fits, low values of mass and radius are
found. The resulting best fits to the bolometric data have slightly
larger (by about 10\%) radius than the fits to the two-band data. The
inferred inclination angle is larger for the bolometric fits while
the spot appears to be closer to the pole in the bolometric fits.

The region of the mass-radius plane consistent with the bolometric
light curve (at the $2 \sigma$ and $3 \sigma$ level) is shown in
Figure~\ref{fig:mr2}. 
Comparison of Figures~\ref{fig:mr1} and
\ref{fig:mr2} shows that the allowed region of the mass-radius
plane for the two-band data is inside of the allowed region for
the bolometric data. 
The range of masses and radii allowed at the $3 \sigma$
level is much wider than the similar allowed region for the two-band
light curves (for oblate stars including time delays) shown in 
Figure~\ref{fig:mr1}. 
This may be due to the larger error bars in the bolometric light curve
compared to the  two-band light curve, or it may be due to the two-band light
curve including a long enough time interval that real variability in the
light curve results in a false bias in the resulting fitted parameters.
More observations of SAX J1808 will be needed to determine which is the
correct case.

The bolometric light curve is consistent with many neutron star
and quark-hadron hybrid stars, as well as pure quark stars,
as can be seen in Figure~\ref{fig:mr2}. 
At the 3 $\sigma$ level, the largest radius star has a radius
of 12.1 km and a mass of $1.2 M_\odot$. The largest mass star allowed
has a radius of 10.6 km and a mass of $1.4 M_\odot$.

\section{Discussion}

In this paper we have revisited light curves constructed from 
SAX~J1808's 1998 outburst. Through our analysis, we
have derived constraints on the mass and radius of the compact
star which depend on the data analysis method and on the selection of
data. In this paper we
investigated the effect on the best-fit mass and radius of:
the inclusion of phase-dependent time-delays; 
the inclusion of oblateness;
the assumed spectral model;
and the binning and selection of data. 

Our results can be summarized as follows. The inclusion of 
phase dependent time-delays and the star's oblate surface both
tend to force the best-fit stellar models to have a smaller radius.
The inclusion of these effects on the mass-radius confidence 
contours causes a significant shrinkage of the allowed values of 
SAX J1808's mass and radius. The magnitude of these changes is
larger than would occur for reasonable changes in the spectral
model. 

We found that the binning and selection of the data has a 
very large effect on the fitted parameter values. The data
published by PG03 corresponds to 19 days of data binned into
two energy bands (3-4 keV and 12-18 keV) where photons with
energies outside of these two bands are excluded from the
analysis. Our analysis of this data agrees with the 
analysis of the same data by PG03 when time-delays are omitted
and a spherical surface is used. When we include the time-delays
and add an oblate surface the allowed values of mass and radius 
are only consistent with quark stars with a very large bag constant.

The compact stars allowed by the two-band data are very
small in both size and mass. If SAX J1808 is described by
such a small star, it would suggest that a phase transition
between hadronic neutron stars and bare quark stars exists.
However, SAX J1808 is accreting matter, so it can't be 
described by a truly bare quark star. We are unaware of any
models for quark stars with accreted hadronic matter that have
stellar parameters falling into the region of the 
mass-radius plane allowed by the two-band data. 

The alternative data selection used by \citet{Pap05} used only
one day of data (in the same 19 day period used by PG03). 
Since SAX J1808's pulse shape was variable during the time period analyzed by PG03 
(see \citet{Har06}), use of data from a single day of observation 
alleviates this significant
problem of mixing data with variable properties.
All photons
in this data set in the range of 2-60 keV were combined into one
bolometric light curve by \citet{Pap05}. We used a similar spectral
model and assumptions to fit the bolometric light curve and found
much less restrictive results. Many neutron star and hybrid
quark-hadron EOS are allowed at the $3 \sigma$ level by fits
to the bolometric data. The largest allowed radius is 12.1 km
(see Figure~\ref{fig:mr2}). 

In all of our models we have restricted our attention to circular
spots with uniform surface brightness. 
More
complicated spot patterns have been predicted by
MHD simulations of accretion \citep{KR05}. However,
at the current level of accuracy in the data
there is no need to consider more complicated spot shapes.
The spots in our models have a constant size (as measured 
at the star's surface). When we varied the spot size, we found
that the changes in the best-fit values of the star's mass
and radius were not very large. However, the best-fit values of
the inclination angle did vary by a fairly large amount. For
this reason we do not quote any best-fit values for the 
inclination angle for SAX J1808.

The two-band data has a couple of features that 
suggest that the more conservative limits set by the bolometric
light curve should be preferred. The two-band data light curves
have omitted the data in the 4 - 12 keV  and 18 - 60 keV ranges. 
This results in
fits that may be skewed in favour of the two energy ranges that 
were selected resulting in very small stars. The bolometric
data includes the photons in the entire range of collected data.
However, the main advantage of the \cite{Pap05} lightcurve is that the data 
is selected from one day only, which avoids the complication of combining variable pulse shapes
over the 19 day period used by PG03.

Our results favour a soft EOS for SAX J1808. 
The 2 $\sigma$ confidence limits for the bolometric
light curve only allow soft EOS which
(with two exceptions) have low maximum masses
below $1.6 M_\odot$. (The exceptions corresponds to quark
star models with low bag constant, such as the Q140 EOS
and the hadronic EOS of the type calculated by \citet{BBB}.)
Our $3 \sigma$ confidence limits
allow stiffer EOS (such as APR) with maximum masses above $2.0 M_\odot$.

In contrast,
recent measurements of the
quiescent flux from \saxj \citep{Heinke}
suggest a stiff EOS due to the very low inferred
luminosity. While the measurements by \citet{Heinke} are quite robust,
further exploration of cooling processes in quark
and other soft EOS are probably required to truly
rule out a soft EOS solely on the basis of
observations during quiescence. 

Another calculation
of the radius based on magnetospheric arguments
\citep{Li}
also suggest a soft EOS. However, \citet{Rap} have
shown that the standard magnetospheric description
of accretion onto a fast pulsar may not hold.

The bolometric results are consistent with a number of other EOS constraints
derived for other neutron stars. A pulse shape analysis for the
slowly rotating X-ray pulsar \herx \citep{Leahy04} also favoured
a softer EOS. The bounds on \herx are more restrictive than ours
in that all EOS allowed (at the 3 $\sigma$ level) by the \herx
analysis are also allowed by our analysis of \saxj. The softest
quark EOS allowed by our analysis is not allowed by the \herx
data.

An analysis of X-ray bursts originating from EXO 0748-676
by \citet{Ozel} predicts a stiff EOS at the 1 $\sigma$ confidence level.
However at the 2 $\sigma$ confidence level the analysis
allows softer EOS (see, for example \citet{Alford}) compatible with
the bolometric results.

The fact that \saxj is an accreting neutron star and has
a very rapid rotation rate suggests that it may have
accreted a large amount of mass. However, it should
be remembered that a high spin rate does not necessarily
require a large accretion of mass. \citet{Cook} showed
that it is possible to spin a $1.4 M_\odot$ star up to
640~Hz by accreting as little as $0.1 M_\odot$
if the EOS is soft. Smaller mass stars have lower
moments of inertia and are easier to spin up, so it is
plausible that the models (allowed by the bolometric
data) with $M=1.3 M_\odot$
could have been born with a mass as high as $1.2 M_\odot$,
consistent with the masses of neutron stars in binary
pulsar systems \citep{Stairs}.

\acknowledgments
This research was supported by grants from NSERC. 
We thank Ingrid Stairs for helpful comments about the 
masses of radio pulsars. We thank the referee for
constructive comments which helped us improve the 
presentation of our results. 
S.~M.~M. thanks the Pacific Institute for Theoretical
Physics and the Department of Physics and Astronomy
at the University of British Columbia for hospitality
during her sabbatical.

\clearpage


\begin{deluxetable}{rrrrrrr}
\tablecaption{Fiducial best-fit stellar models for two-band data.
Models include exact light-bending,
no time-delays and a spherical surface. \label{fid}}
\tablewidth{0pt}
\tablehead{
\colhead{$2M/R$}&\colhead{$M$}&\colhead{$R$}&\colhead{$\theta$}&
\colhead{$i$}&\colhead{$a$}&\colhead{$\chi^2/$dof}\\
\colhead{}&\colhead{$M_\odot$}&\colhead{km}&\colhead{deg.}&
\colhead{deg.}&\colhead{}&\colhead{}
}
\startdata
0.60 & 1.07 & 5.30 & 20.5 & 63.5 & 0.603 & 31.1/24 \\ 
0.50 & 1.03 & 6.10 & 22.9 & 44.6 & 0.557 & 35.1/24 \\ 
0.40 & 1.00 & 7.38 & 18.6 & 44.7 & 0.570 & 38.6/24 \\ 
0.30 & 0.83 & 8.13 & 22.7 & 31.5 & 0.538 & 40.9/24 \\ 
0.20 & 0.70 & 10.29 & 13.8 & 42.3 & 0.589 & 43.5/24 
\enddata
\end{deluxetable}

\begin{deluxetable}{rrrrrrr}
\tablecaption{Best-fit stellar models for two-band data.
Models include approximate light-bending,
no time-delays and a spherical surface.
\label{tab:approx}}
\tablewidth{0pt}
\tablehead{
\colhead{$2M/R$}&\colhead{$M$}&\colhead{$R$}&\colhead{$\theta$}&
\colhead{$i$}&\colhead{$a$}&\colhead{$\chi^2/$dof}\\
\colhead{}&\colhead{$M_\odot$}&\colhead{km}&\colhead{deg.}&
\colhead{deg.}&\colhead{}&\colhead{}
}
\startdata
0.60 & 1.16 & 5.69 & 16.6 & 77.6 & 0.671 & 32.6/24 \\ 
0.50 & 1.03 & 6.08 & 23.9 & 42.7 & 0.552 & 35.1/24 \\ 
0.40 & 0.95 & 7.04 & 24.4 & 34.5 & 0.540 & 38.3/24 \\ 
0.30 & 0.88 & 8.67 & 18.7 & 36.7 & 0.551 & 42.9/24 \\ 
0.20 & 0.64 & 9.42 & 19.1 & 32.1 & 0.544 & 42.8/24
\enddata
\end{deluxetable}

\begin{deluxetable}{lrrrrrrr}
\tablecaption{Model dependence of two-band fits.\label{tab:spectra}}
\tablewidth{0pt}
\tablehead{
\colhead{Model} &
\colhead{$2M/R$}&\colhead{$M$}&\colhead{$R$}&\colhead{$\theta$}&
\colhead{$i$}&\colhead{$a$}&\colhead{$\chi^2/$dof}\\
\colhead{}&\colhead{}&\colhead{$M_\odot$}&\colhead{km}&\colhead{deg.}&
\colhead{deg.}&\colhead{}&\colhead{}
}
\startdata
Fiducial & 0.60 & 1.07 & 5.30 & 20.5 & 63.5 & 0.603 & 31.1/24 \\ 
$\Gamma = 1.8$ & 0.60 & 1.08 & 5.33 & 22.0 & 59.7 & 0.590 & 35.9/24 \\ 
$\tau = 0.1$ & 0.60 & 1.09 & 5.37 & 20.4 & 62.6 & 0.596 & 29.3/24 \\
$b$ free & 0.60 & 1.06 & 5.23 & 21.0 & 61.9 & 0.595 & 26.9/24 \\ 
$r_{sp} = 1 \km$ & 0.60 & 1.06 & 5.20 & 20.1 & 61.8 & 0.591 & 30.9/24 \\ 
$r_{sp} = 2 \km$ & 0.60 & 1.07 & 5.19 & 21.5 & 62.9 & 0.601 & 34.1/24 \\
$r_{sp} = 3 \km$ & 0.60 & 1.04 & 5.13 & 27.3 & 50.8 & 0.574 & 35.7/24 
\enddata
\end{deluxetable}

\begin{deluxetable}{rrrrrrr}
\tablecaption{Best-fit stellar models for two-band data
 including time-delays.\label{tab:td}
}
\tablewidth{0pt}
\tablehead{
\colhead{$2M/R$}&\colhead{$M$}&\colhead{$R$}&\colhead{$\theta$}&
\colhead{$i$}&\colhead{$a$}&\colhead{$\chi^2$}\\
\colhead{}&\colhead{$M_\odot$}&\colhead{km}&\colhead{deg.}&
\colhead{deg.}&\colhead{}&\colhead{}
}
\startdata
0.60 & 0.98 & 4.81 & 28.4 & 46.3 & 0.548 & 36.6/24 \\ 
0.50 & 0.98 & 5.78 & 33.1 & 30.8 & 0.534 & 41.9/24 \\ 
0.40 & 0.93 & 6.88 & 28.0 & 30.0 & 0.533 & 45.7/24 \\ 
0.30 & 0.81 & 7.97 & 27.4 & 25.9 & 0.531 & 48.7/24 \\ 
0.20 & 0.63 & 9.26 & 19.8 & 30.9 & 0.538 & 50.7/24
\enddata
\end{deluxetable}

\begin{deluxetable}{rrrrrrr}
\tablecaption{Best-fit stellar models for two-band data using exact light-bending,
time-delays and oblate surface. \label{tab:ob}}
\tablewidth{0pt}
\tablehead{
\colhead{$2M/R$}&\colhead{$M$}&\colhead{$R$}&\colhead{$\theta$}&
\colhead{$i$}&\colhead{$a$}&\colhead{$\chi^2/$dof}\\
\colhead{}&\colhead{$M_\odot$}&\colhead{km}&\colhead{deg.}&
\colhead{deg.}&\colhead{}&\colhead{}
}
\startdata
0.60 &0.96 &4.72 &31.0 &42.9 &0.542 &36.3/24 \\
0.50 &0.97 &5.75 &26.2 &40.2 &0.543 &41.6/24 \\
0.40 &0.90 &6.68 &30.2 &28.8 &0.532 &45.5/24 \\
0.30 &0.77 &7.63 &25.4 &29.8 &0.533 &48.4/24 \\
0.20 &0.56 &8.37 &24.3 &28.5 &0.534 &50.7/24 \\
\enddata
\end{deluxetable}

\begin{deluxetable}{rrrrrrr}
\tablecaption{Best-fit models for bolometric light curves
using a spherical surface and no time delays.
\label{tab:Pap-notd}}
\tablewidth{0pt}
\tablehead{
\colhead{$2M/R$}&\colhead{$M$}&\colhead{$R$}&\colhead{$\theta$}&
\colhead{$i$}&\colhead{$a$}&\colhead{$\chi^2/$dof}\\
\colhead{}&\colhead{$M_\odot$}&\colhead{km}&\colhead{deg.}&
\colhead{deg.}&\colhead{}&\colhead{}
}
\startdata
0.60 & 0.98 & 4.75 & 19.5 & 63.9 & 0.604 & 20.4/58 \\  
0.50 & 0.96 & 5.65 & 18.1 & 53.5 & 0.593 & 21.0/58 \\ 
0.40 & 0.89 & 6.59 & 17.0 & 45.9 & 0.580 & 21.7/58 \\ 
0.30 & 0.77 & 7.62 & 16.2 & 40.9 & 0.570 & 20.9/58 \\ 
0.20 & 0.58 & 8.57 & 15.5 & 36.8 & 0.564 & 21.4/58 \\ 
\enddata
\end{deluxetable}

\begin{deluxetable}{lrrrrrrr}
\tablecaption{Model Dependence of Fits to Bolometric Data.
Fits are computed using exact light-bending, no time delays,
and a spherical surface.\label{tab:Pap-models}}
\tablewidth{0pt}
\tablehead{
\colhead{Model} &
\colhead{$2M/R$}&\colhead{$M$}&\colhead{$R$}&\colhead{$\theta$}&
\colhead{$i$}&\colhead{$a$}&\colhead{$\chi^2/$dof}\\
\colhead{}&\colhead{}&\colhead{$M_\odot$}&\colhead{km}&\colhead{deg.}&
\colhead{deg.}&\colhead{}&\colhead{}
}
\startdata
$r_{sp}=1.5 \km$ & 0.60 & 0.98 & 4.75 & 19.5 & 63.9 & 0.604 & 20.4/58 \\ 
$r_{sp}=1.0 \km$ & 0.60 & 0.94 & 4.60 & 19.0 & 63.5 & 0.603 & 20.1/58 \\  
$r_{sp}=2.0 \km$ & 0.60 & 0.93 & 4.59 & 24.1 & 53.0 & 0.569 & 20.6/58\\ 
$r_{sp}=3.0 \km$ & 0.60 & 0.97 & 4.79 & 24.2 & 50.6 & 0.565 & 21.7/58\\
$\Gamma=1.8$ & 0.60 & 0.99 & 4.89 & 19.5 & 63.7 & 0.605 & 20.4/58 \\ 
\enddata
\end{deluxetable}

\begin{deluxetable}{rrrrrrr}
\tablecaption{Best-fit models for bolometric light curves.
Models include exact light-bending,
time-delays and oblate surface.  
\label{tab:Pap-ob}}
\tablewidth{0pt}
\tablehead{
\colhead{$2M/R$}&\colhead{$M$}&\colhead{$R$}&\colhead{$\theta$}&
\colhead{$i$}&\colhead{$a$}&\colhead{$\chi^2/$dof}\\
\colhead{}&\colhead{$M_\odot$}&\colhead{km}&\colhead{deg.}&
\colhead{deg.}&\colhead{}&\colhead{}
}
\startdata
0.60 & 0.99 & 4.87 & 17.7 & 70.7 & 0.630 & 21.7/58 \\
0.50 & 1.05 & 6.21 & 13.9 & 68.4 & 0.667 & 22.1/58 \\
0.40 & 1.04 & 7.73 & 11.2 & 67.4 & 0.720 & 22.6/58 \\
0.30 & 0.85 & 8.56 & 10.8 & 61.0 & 0.714 & 23.0/58 \\
0.20 & 0.60 & 9.17 & 10.6 & 57.4 & 0.727 & 23.5/58 \\
\enddata
\end{deluxetable}

\clearpage


\end{document}